%% file: note.tex
\documentclass[11pt]{article}
\usepackage{graphicx}
\usepackage{multirow}
\usepackage{dcolumn}
\usepackage{bm}
\usepackage{lineno}

\newcommand{\BABARPubYear}    {07}
\newcommand{\BABARConfNumber}  {005}
\newcommand{\SLACPubNumber} {12732}

\input babarsym

\input privsym

\long\def\inst#1{\par\nobreak\kern 4pt\nobreak
    {\it #1}\par\vskip 10pt plus 3pt minus 3pt}

\begin{document}

{\pagestyle{empty}

\begin{flushright}
\babar-CONF-\BABARPubYear/\BABARConfNumber \\
SLAC-PUB-\SLACPubNumber \\
August 2007 \\
\end{flushright}

\par\vskip 5cm

\begin{center}
\Large \bf Measurement of The Time-Dependent \CP Asymmetry in $\Bz \to \Kstarz \gamma$ Decays
\end{center}
\bigskip

\begin{center}
\large The \babar\ Collaboration\\
\mbox{ }\\
\today
\end{center}
\bigskip \bigskip

\begin{center}
\large \bf Abstract
\end{center}
\input abstract.tex

\vfill
\begin{center}
Contributed to the
XXIII$^{\rm rd}$ International Symposium on Lepton and Photon Interactions at High~Energies, 8/13 -- 8/18/2007, Daegu, Korea
\end{center}

\vspace{1.0cm}
\begin{center}
{\em Stanford Linear Accelerator Center, Stanford University, 
Stanford, CA 94309} \\ \vspace{0.1cm}\hrule\vspace{0.1cm}
Work supported in part by Department of Energy contract DE-AC02-76SF00515.
\end{center}

\newpage
} 

\input authors_lp2007.tex

The radiative decay \btosgam~\cite{ref:cc} serves as a probe of
physics beyond the standard model (SM)~\cite{Atwood:1997zr}. In the SM
it proceeds at leading order through a loop diagram involving a
virtual $W^-$ boson and $t$ quark. Because weak interactions involve
only left-handed fermions and right-handed anti-fermions, the photon
in \btosgam is predominantly left-handed while in $\overline{b} \to
\overline{s} \gamma$ it is right-handed. Any possible interference
between the direct decay $\Btokstcpg$ and the decay via \Bz mixing,
$\Bz \to \Bzb \to \Kstarz \gamma$, is suppressed by the small rate of 
$b \to s \gamma_R$ relative to $b \to s \gamma_L$, which is of the
order $m_s/m_b$. SM predictions of the \CP asymmetry due to
interference between mixing and decay are expected
to be about $-0.02$~\cite{Ball:2006cv}. As discussed
in Ref.~\cite{Atwood:1997zr}, left-right symmetric
models~\cite{Mohapatra:1974hk} could conceivably produce
mixing-induced \CP asymmetries larger than $0.5$. Some supersymmetric
models without left-right symmetry~\cite{Chun:2000hj} have also been shown to
permit \CP asymmetries of $\order(1)$. Because the SM asymmetry is
quite small, any significant evidence of a larger asymmetry would
point to a source beyond the SM.

Here we report a preliminary updated measurement of the time-dependent
\CP asymmetry in \Btokstg\ based on $431 \times
10^6$ $\FourS \to \BB$ decays collected by the \babar\ detector at the
PEP-II asymmetric-energy \epem collider at SLAC. Previous measurements
of this mode were performed by \babar~\cite{Aubert:2005bu} and
Belle~\cite{Ushiroda:2006fi}.

The \babar\ detector is described in detail
elsewhere~\cite{Aubert:2001tu}. Most important 
to this analysis are the five-layer, double-sided silicon microstrip
detector (SVT), the 40-layer drift chamber, and the CsI(Tl) electromagnetic
calorimeter (EMC). A detailed Monte Carlo (MC) simulation of signal and
background processes was performed using the {\tt EVTGEN}~\cite{Lange:2001uf}
generator and the {\tt GEANT4} package~\cite{Agostinelli:2002hh}.

Time-dependent \CP asymmetries are determined using the difference of
\Bz meson proper decay times $\deltat \equiv t_{\rm sig} - t_{\rm tag}$,
where $t_{\rm sig}$ is the proper decay time of the signal $B$ ($B_{\rm sig}$)
and $t_{\rm tag}$ is that of the other $B$ ($B_{\rm tag}$). The
\deltat\ for $B_{\rm sig}$ decaying to a final state $f$ is
distributed according to 
\begin{equation}
{\cal{P}}_\pm (\Delta t) = \frac{e^{-|\Delta t|/\tau_B}}{4 \tau_B} \times
\left[1 \pm S_f \sin(\Delta m_d \Delta t) \mp C_f \cos(\Delta m_d \Delta t)
\right],
\label{eq:deltatpdf} \end{equation}
where the upper and lower signs correspond to the tag-side $B$ having
flavor \Bz and \Bzb respectively, $\tau_B$ is the \Bz lifetime, and
$\Delta m_d$ is the 
$\Bz - \Bzb$ mixing frequency. The $C_f$ coefficient is associated with
the difference in decay amplitudes for $\Bz \to f$ and $\Bzb \to f$,
while the $S_f$ coefficient involves interference between the $\Bz -
\Bzb$ mixing and decay amplitudes. We note that direct \CP violation
in \Btokstg\ decays is predicted to be smaller than 1\% in the
SM~\cite{Kagan:1998bh}. The current evidence is consistent with this,
based on self-tagging $B \to K^* \gamma$ decays~\cite{Barberio:2007cr}.

We search for \Btokstcpg\ candidates based on the following
criteria, all but one of which were used in the previous result.
Photon candidates are required to have energy greater 
than 30~\mev and must have the expected shower shapes in the EMC. The
photon from the $B$ decay, also called the primary photon, is required
to have an energy between 1.5 and 3.5~\gev in the $\epem$ center of
mass (CM) frame to be consistent with \btosgam
decays~\cite{Aubert:2006gg}. It must be isolated from other charged and neutral
clusters in the EMC. Primary photon candidates that form $\piz \to
\gaga$ or $\eta \to \gaga$ candidates of invariant mass $115 <
m_{\gaga} < 155\mevcc$ or $470 < m_{\gaga} < 620\mevcc$, respectively,
when combined with another photon of energy greater than 50~\mev for
\piz or 250~\mev for $\eta$ are
discarded. We select $\KS \to \pip \pim$ candidates from oppositely
charged tracks for which the confidence level of the vertex fit is
greater than 0.1\%, the
$\pip\pim$ invariant mass is between 487 and 508~\mevcc (about
$3\sigma$), and the reconstructed decay length is greater than 5 times its
uncertainty. We select $\piz \to \gaga$ candidates with invariant mass
between 115 and 155~\mevcc (about $3\sigma$) and energy greater than
590~\mev in the lab frame. 
We require the invariant $\KS \piz$  mass \mkspiz\ to be within $0.8 -
1.0 \gevcc$, and later
use its shape in a maximum likelihood fit. We require
$|\cos{\theta_{K^*}}| < 0.9$, where
$\theta_{K^*}$ is the angle between the \KS and the primary photon in
the $\KS \piz$ rest frame. Along with signal candidates, we also
reconstruct $B^+ \to K^{*+}(\KS \pip) \gamma$ candidates subject to
the same requirements as \Bz candidates, and veto
events for which the invariant $\KS \pip$ mass is within $0.8 - 1.0
\gevcc$. This is new since the last result, and it removes 12\% of
the background due to non-signal $B$ decays.

We identify signal decays using two Lorentz-invariant quantities: the
energy-substituted mass $\mes = \left(\sqrt{ (s/2 + c^2 {\bf p}_{\epem}
\cdot {\bf p}_B)^2/E^2_{\epem} - |p_B|^2}\right)/c^2$ and the energy
difference $\DeltaE = E^*_B - \sqrt{s}/2$,
where $(E_{\epem},c {\bf p}_{\epem})$ and $p_B \equiv (E_B, c {\bf
p}_B)$ are the four-momenta of
the initial \epem system and the $B$ candidate, respectively,
$\sqrt{s}$ is the CM energy, and the asterisk denotes the
CM frame. We require $5.2 < \mes < 5.3\gevcc$ and $|\DeltaE| < 250\mev$. 
To discriminate $B$ decays against continuum $\epem~\to~\qqbar~(q~=~u,d,s,c)$
background we require $|\cos{\theta^*_B}| < 0.9$, 
where $\theta^*_B$ is the CM angle between the $B$ candidate and the
$e^-$ beam direction. We also exploit event topology by
requiring the ratio of Legendre moments \leg\ to be less than 0.55,
where $L_i = \sum_j |p^*_j||\cos{\theta^*_j}|^i$, $p^*_j$ is the
CM momentum of each particle $j$ not used to reconstruct the $B$
candidate, and $\theta^*_j$ is the CM angle between $p^*_j$ and the
thrust axis of the reconstructed $B$ candidate.

After all selection criteria have been applied we find the average
candidate multiplicity in events with at least one candidate is
1.15. In these cases we select the candidate with \piz mass closest to
its nominal value~\cite{Yao:2006px}, and if there is an ambiguity then
we select the one with the \KS mass closest to its nominal value.
We evaluate the selection efficiency using
simulated events. We find it is about 16\%, and combined with the 
$\Bz \to \Kstarz \gamma$ branching fraction, $\BR(\Bz \to \Kstarz
\gamma) = (4.01 \pm 0.20) \times 10^{-5}$~\cite{Barberio:2007cr}, we
expect $312 \pm 24$ signal events.
We also expect approximately 35 events originating from non-signal $B$
decays (\BB background). The rest of the 3677 selected events come
from continuum background. These two background types are treated
separately below.

For each reconstructed \Btokstg\ candidate we use the remaining tracks in
the event to determine the decay vertex position and flavor of $B_{\rm tag}$.
A neural network based on kinematic and particle identification
information assigns each event to one of seven mutually exclusive
tagging categories~\cite{Aubert:2004zt}, including a category for
events in which a flavor tag is not determined. The performance of this
algorithm is determined
using a data sample ($B_{\rm flav}$ sample) of fully-reconstructed
$\Bz \to D^{(*)-} \pip / \rho^+/a^+_1$ decays. The average tagging
efficiency is measured to be
$Q=\sum_c\epsilon^c(1-2w^c)^2=(31.2\pm0.3)\%$, where $\epsilon^c$ and
$w^c$ are the efficiencies and mistag probabilities, respectively, for
events tagged in category $c$.

We determine the proper time difference between $B_{\rm sig}$ and 
$B_{\rm tag}$ from the spatial separation between their decay
vertices in the same way as our previous measurement. The $B_{\rm sig}$
vertex is reconstructed by combining the \KS 
trajectory with the knowledge of the average interaction point (IP), which
is calculated every ten minutes based on two-track events during
data-taking. The $B_{\rm tag}$ vertex is reconstructed from the
remaining charged tracks in the event~\cite{Aubert:2002rg}. We compute
\deltat and its uncertainty from a geometric
fit~\cite{Hulsbergen:2005pu} to the $\FourS \to \Bz\Bzb$ system, which
takes the IP constraint~\cite{Aubert:2004pe} into account. The
resolution of \deltat is improved by constraining the average sum of the two
$B$ decay times $(t_{\rm sig} + t_{\rm tag})$ to equal $2 \tau_{\Bz}$,
with an uncertainty of $\sqrt{2}\tau_{\Bz}$. We have verified in
signal MC that no bias on $S_{K^* \gamma}$ or $C_{K^* \gamma}$ results
from this procedure.

The \deltat resolution strongly depends on the number of SVT layers
traversed by the pions from the \KS. In order for the \deltat
information to be useful, we require that each pion
have at least 2 hits in the SVT, and that $\sigma_{\deltat} < 2.5$~ps and
$|\deltat| < 20$~ps. About 70\% of the events in the data sample pass these
requirements. The events for which the \deltat information is not used
can still contribute to the measurement of the $C_{K^* \gamma}$
parameter as long as they have flavor tagging information.

We extract signal yields and \CP asymmetries using an unbinned maximum
likelihood fit to \mes, \DeltaE, \leg, \mkspiz, flavor tag, \deltat, and
$\sigma_{\deltat}$. 
As stated earlier, we
expect a significant contribution from \BB background, so
we extract the event yield from this source as well as
continuum background. The likelihood
function is the same one used in the previous version of this
analysis, and is described in detail in Ref.~\cite{Aubert:2004pe}. We
assume that the correlation among the observables is
small enough that the likelihood function can be constructed as a
product of one-dimensional probability density functions (PDF). A
systematic correction is applied later as a result of this
assumption. All signal PDF
parameters are determined using simulated events, except for the
flavor tag efficiencies, mistag probabilities, and \deltat resolution
function parameters, which are determined from the $B_{\rm flav}$
sample. \BB background shapes are also determined from simulation. We
use the large fraction of background events in the fitted data sample
to determine continuum background PDF parameters.

The \deltat PDF for signal events and \BB background is obtained from
the convolution of Eq.~\ref{eq:deltatpdf} with a resolution function
${\cal{R}} (\delta t \equiv (\deltat - \deltat_{\rm true}),
\sigma_{\deltat})$. The \CP asymmetries for the \BB background, $S_{\BB}$ and
$C_{\BB}$, are fixed to zero in the fit, and we account for a 
possible deviation from zero in the systematic uncertainty. The
resolution function is parameterized as the sum of three gaussian
distributions~\cite{Aubert:2002rg}. The first two have a nonzero mean
proportional to the reconstructed $\sigma_{\deltat}$, accounting for a
small bias in \deltat from charm decays of the $B_{\rm tag}$. Their
width is also proportional to $\sigma_{\deltat}$. The third gaussian is
centered at zero with a fixed width of 8 ps. We have verified in
simulation that the parameters of the resolution function for
\Btokstg\ events are compatible with those obtained from the $B_{\rm flav}$
sample. Therefore we use the $B_{\rm flav}$ parameters for better
precision. We assume that the continuum background contains only
prompt decays and find that the \deltat distribution is well-described
by a resolution function of the same form used by the signal PDF. The
parameters of the background resolution function are determined in the
fit to data.

\begin{figure}
\begin{center}
    \includegraphics[width=0.45\linewidth]{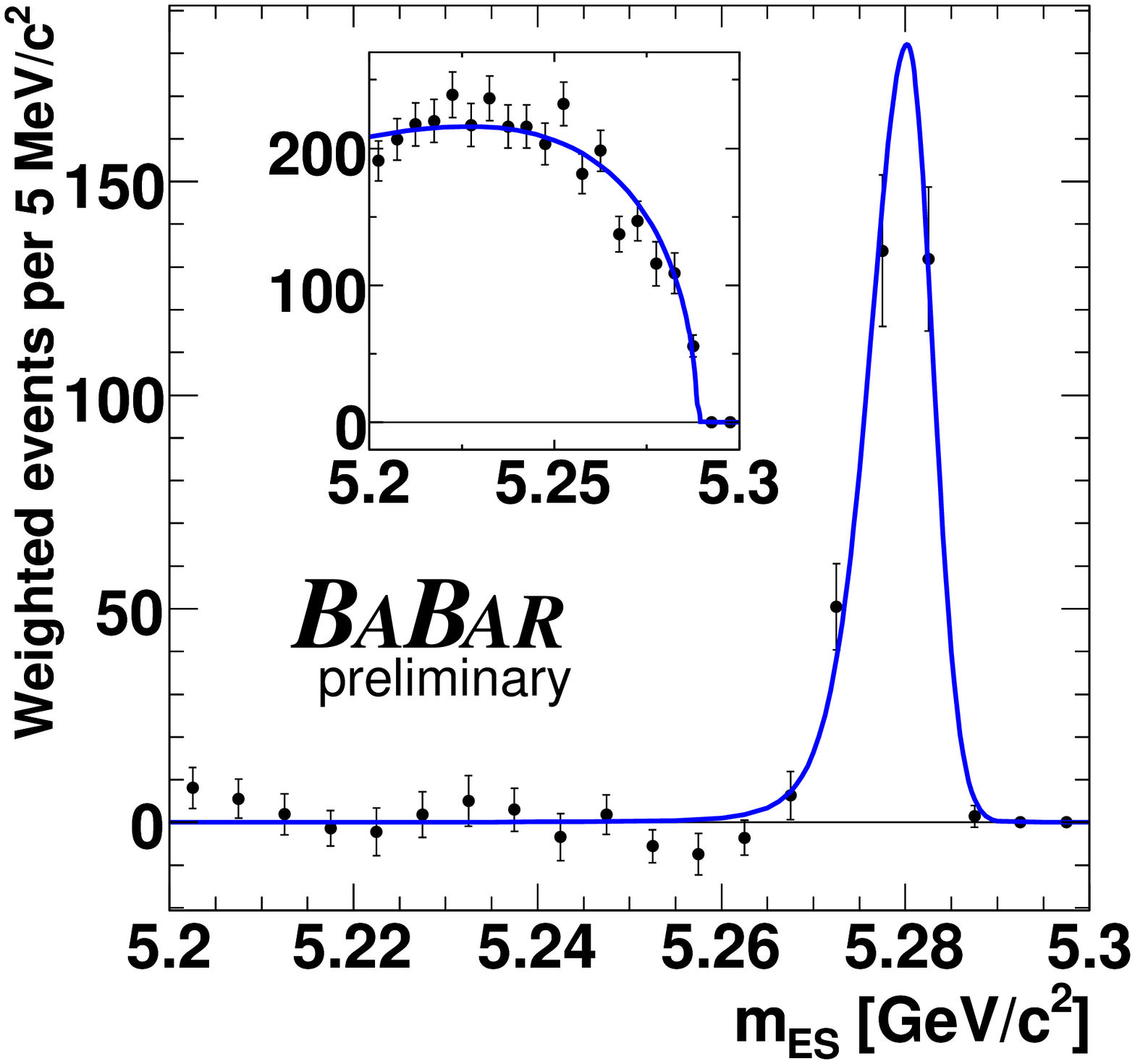}
    \includegraphics[width=0.45\linewidth]{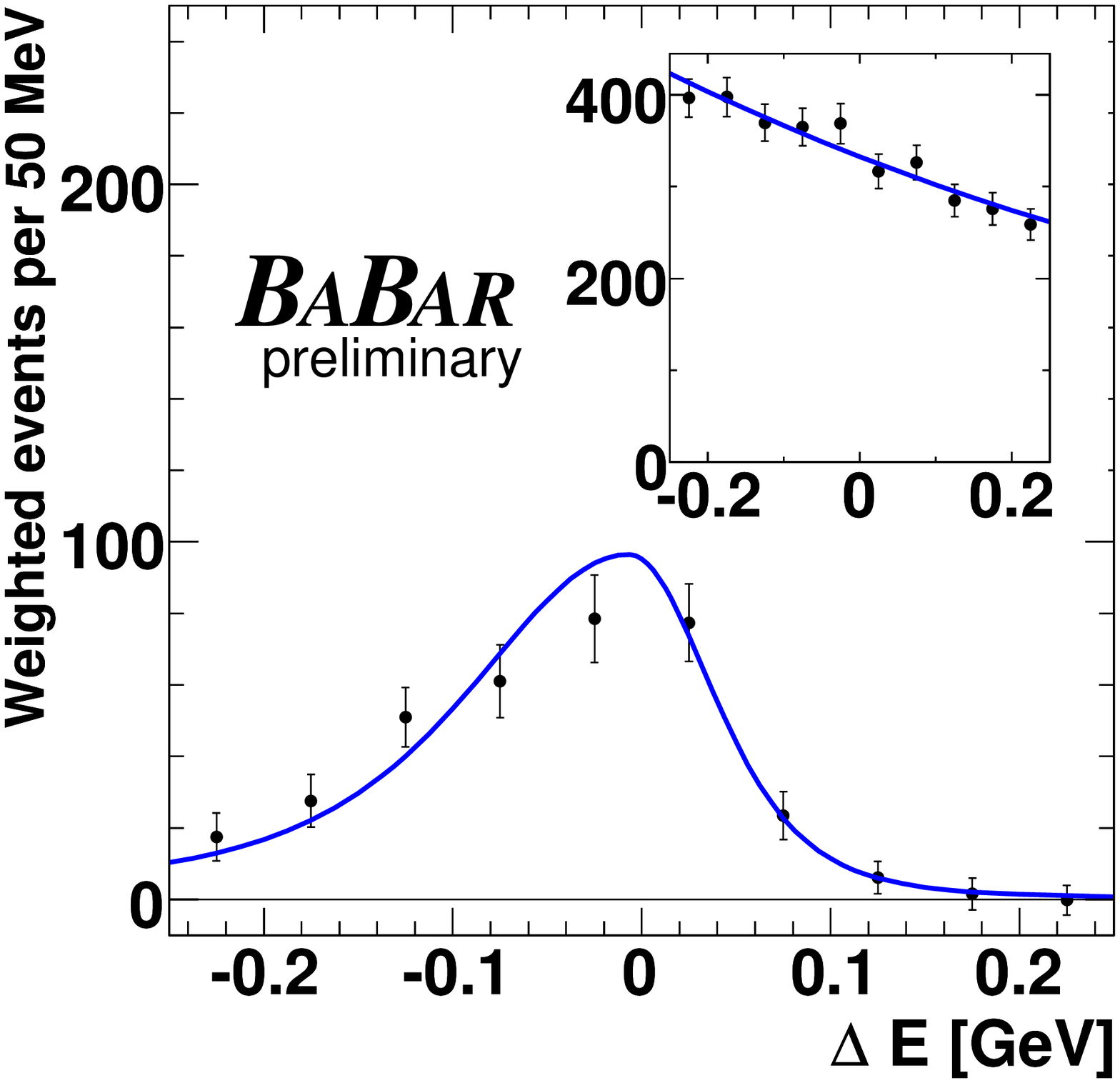}
\end{center}
  \caption{ Signal and background (inset) distributions for \mes{}
(left) and \DeltaE{} (right)
obtained with the weighting technique described in
Ref.~\cite{Pivk:2004ty}.  The curves
represent the PDFs used in the fit, normalized to the fitted yield.}
  \label{fig:kstgMesDe}
\end{figure}

\begin{figure}
  \centerline{\includegraphics[width=0.45\linewidth]{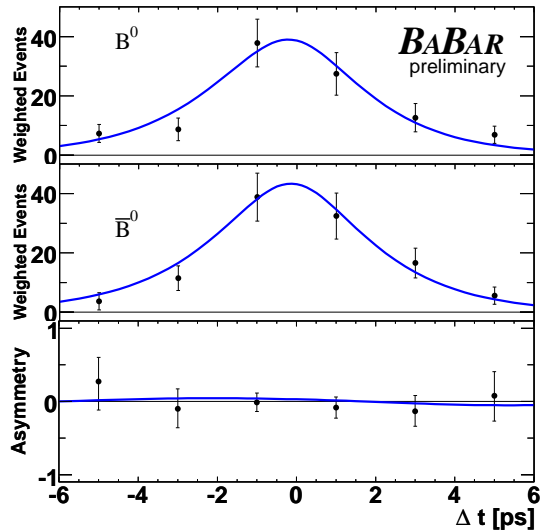}}
  \caption{Signal distribution for $\deltat$ 
obtained with the weighting technique
described in Ref.~\cite{Pivk:2004ty}, with $B_{\rm tag}$ tagged as
$\Bz$ (top) or
$\Bzb$ (center), and the asymmetry (bottom). The curves represent the
PDFs for signal decays in the likelihood fit, normalized to the final
fit result.}
  \label{fig:kstgDtAsym}
\end{figure}

Figure~\ref{fig:kstgMesDe} shows the background-subtracted
distributions for \mes\ and \DeltaE\ for \Btokstg\ candidates. The
background subtraction is performed using the sPlot event weighting
technique described in Ref.~\cite{Pivk:2004ty}. The curves in the figure
represent the signal PDFs used in the fit. Figure~\ref{fig:kstgDtAsym}
shows the background-subtracted distributions of \deltat for \Bz- and
\Bzb-tagged events, and the asymmetry as a function \deltat.

We find $316 \pm 22$ signal events
with \[S_{K^* \gamma} = -0.08 \pm 0.31 \pm 0.05\] and 
\[C_{K^* \gamma} = -0.15
\pm 0.17 \pm 0.03,\] where the first error is statistical and the second
systematic. We discuss systematic uncertainties below. The statistical
uncertainties have been increased beyond what was reported in the fit
result because we have determined them to
be underestimated, using an ensemble of simulated experiments in which
events were generated from the likelihood PDFs. The scaling factors for
$S_{K^* \gamma}$ and $C_{K^* \gamma}$ are 1.097 and 1.035
respectively. Because the 
uncertainty of $C_{K^* \gamma}$ is larger than that obtained from the
partial rate asymmetry in self-tagging $B \to K^* \gamma$
decays~\cite{Barberio:2007cr}, we also perform the fit with $C_{K^*
\gamma}$ fixed to zero and find \[S_{K^* \gamma}(C_{K^* \gamma} = 0) =
-0.07 \pm 0.32 \pm 0.05.\] The linear correlation coefficient between
$S_{K^* \gamma}$ and $C_{K^* \gamma}$ is 0.046.

We consider several sources of systematic uncertainties related to the
level and possible asymmetry of the background contribution from other
$B$ decays.  We evaluate this contribution using simulated samples of
non-signal $\B\to\X_s\gamma$ and other $B$ decays. For the
former we use the Kagan-Neubert model~\cite{Kagan:1998ym} to model the
photon energy spectrum and JETSET for the fragmentation of the $s$
quark. Since the final state multiplicity predicted by the
fragmentation model is significantly different from \babar's
measurement~\cite{Aubert:2005cu}, we reweight events
according to their multiplicity. From these studies we expect to find about
$35$ 
\BB events,
with approximately equal contributions from $\B\to\X_s\gamma$ decays
and other generic $B$ decays. The \BB background yield extracted in
the fit to the data is
$22 \pm 22$
events. We vary $S_{\BB}$ and
$C_{\BB}$ within a conservative range derived from the
composition of the \BB background sample and the \CP asymmetry
averages reported by the Heavy Flavor Averaging
Group~\cite{Barberio:2007cr} to assign a systematic
uncertainty due to the assumption of zero asymmetry in this
source. Because the \BB yield in data is smaller than expected,
we fix it to the expected value when we vary its \CP asymmetry within
the $S_{\BB}$ range $\pm0.41$ and the $C_{\BB}$ range $\pm0.33$. We
assign uncertainties of $0.029$ on $S_{K^* \gamma}$ and 
$0.018$ on $C_{K^* \gamma}$ based on these variations.

In an ensemble of simulated experiments created by
generating background events from the PDFs and embedding signal events
from the full MC simulation, we determined there was bias in $S_{K^*
\gamma}$ but not in $C_{K^* \gamma}$.
We apply a correction of $+0.067$ to $S_{K^* \gamma}$, with a systematic
uncertainty of half the shift. We also assign a systematic uncertainty
of 0.015 to $C_{K^* \gamma}$
based on these tests.

Systematic effects due to uncertainties in the resolution function,
beam spot position, and possible SVT misalignment are quantified in
the same manner as Ref.~\cite{newkspi0}. We also include uncertainties
due to imperfect knowledge of the fixed PDF parameters and shapes used in the
fit, amounting to
$0.025$ on $S_{K^* \gamma}$ and $0.010$ on $C_{K^* \gamma}$.
Finally, uncertainties of 0.001 in $S_{K^* \gamma}$ and 0.015 in
$C_{K^* \gamma}$
are included to account for doubly-Cabibbo-suppressed (DCS) decays 
of the $B_{\rm tag}$~\cite{Aubert:2007hm}. The systematic
uncertainties are summarized in Table~\ref{tab:systematics}.

\begin{table}
  \caption{ \Btokstg\ systematic uncertainties. \label{tab:systematics} }
\centerline{
    \begin{tabular}{|c|c|c|}
      \hline\hline
        Source & $\Delta S$ & $\Delta C$ \\
      \hline
        \BB Background               & 0.029 & 0.018 \\
        Bias Uncertainty             & 0.034 & 0.015 \\
        \deltat\ Resolution Function & 0.011 & 0.018 \\
        Beamspot                     & 0.004 & 0.001\\
        SVT Alignment                & 0.002 & 0.001 \\
        PDF Uncertainty              & 0.025 & 0.010 \\
        DCS $B_{\rm tag}$ Decays     & 0.001 & 0.015 \\
        \hline
        Total                        & 0.052 & 0.035 \\
\hline \hline
\end{tabular}
}
\end{table}

In summary we have performed a new preliminary measurement of the
time-dependent \CP asymmetry in \Btokstcpg\ decays. We have found it to be
consistent with our previous result, as well as with the standard model
expectation.

\input acknowledgements

\end{document}

%% file: privsym.tex
\newcommand{\Btokstg}{\ensuremath{\Bz\to\Kstarz\gamma}}
\newcommand{\Btokstcpg}{\ensuremath{\Bz\to\Kstarz(\KS\piz)\gamma}}

\newcommand{\leg}{\ensuremath{L_2/L_0}}
\newcommand{\mkspiz}{\ensuremath{m(\KS \piz)}}

\newcommand{\bi}{\begin{itemize}}
\newcommand{\ei}{\end{itemize}}

%% file: abstract.tex
We present a preliminary measurement of the time-dependent $CP$ asymmetry
in $B^0 \rightarrow K^{*0} (K_S^0 \pi^0) \gamma$ decays based on $431
\times 10^6$ $\Upsilon(4S) \rightarrow B\overline{B}$
decays collected with the \mbox{\slshape B\kern-0.1em{\smaller
A}\kern-0.1em B\kern-0.1em{\smaller A\kern-0.2em R}} detector at the PEP-II
asymmetric-energy $e^+e^-$ collider at SLAC. In a sample containing $316
\pm 22$ signal events we measure $S_{K^{*} \gamma} = -0.08 \pm
0.31 \pm 0.05$ and $C_{K^{*} \gamma} = -0.15 \pm 0.17 \pm
0.03$. 
The uncertainties are statistical and systematic,
respectively.

%% file: authors_lp2007.tex
\begin{center}
\small

The \babar\ Collaboration,
\bigskip

{B.~Aubert,}
{M.~Bona,}
{D.~Boutigny,}
{Y.~Karyotakis,}
{J.~P.~Lees,}
{V.~Poireau,}
{X.~Prudent,}
{V.~Tisserand,}
{A.~Zghiche}
\inst{Laboratoire de Physique des Particules, IN2P3/CNRS et Universit\'e de Savoie, F-74941 Annecy-Le-Vieux, France }
{J.~Garra~Tico,}
{E.~Grauges}
\inst{Universitat de Barcelona, Facultat de Fisica, Departament ECM, E-08028 Barcelona, Spain }
% ...

\end{center}\newpage

%% file: acknowledgements.tex
We are grateful for the 
extraordinary contributions of our \pep2\ colleagues in
achieving the excellent luminosity and machine conditions
that have made this work possible.
The success of this project also relies critically on the 
expertise and dedication of the computing organizations that 
support \babar.
The collaborating institutions wish to thank 
SLAC for its support and the kind hospitality extended to them. 
This work is supported by the
US Department of Energy
and National Science Foundation, the
Natural Sciences and Engineering Research Council (Canada),
the Commissariat \`a l'Energie Atomique and
Institut National de Physique Nucl\'eaire et de Physique des Particules
(France), the
Bundesministerium f\"ur Bildung und Forschung and
Deutsche Forschungsgemeinschaft
(Germany), the
Istituto Nazionale di Fisica Nucleare (Italy),
the Foundation for Fundamental Research on Matter (The Netherlands),
the Research Council of Norway, the
Ministry of Science and Technology of the Russian Federation, 
Ministerio de Educaci\'on y Ciencia (Spain), and the
Science and Technology Facilities Council (United Kingdom).
Individuals have received support from 
the Marie-Curie IEF program (European Union) and
the A. P. Sloan Foundation.

%% file: note.bbl
\begin{thebibliography}{99}
\bibitem{ref:cc}
Unless explicitly stated, charge conjugate decay modes are included
implicitly throughout this paper.
\end{thebibliography}
